\documentclass[prd, twocolumn, nofootinbib, floatfix]{revtex4-1}
\pdfoutput=1

\usepackage{hyperref}
\usepackage{amsmath}
\usepackage{graphicx}
\usepackage{dcolumn}
\usepackage{bm}
\usepackage{epsfig}
\usepackage{amssymb,latexsym,mathrsfs}
\usepackage{graphicx}
\usepackage{color}

\hypersetup{
    colorlinks=true,
    linkcolor=red,
    citecolor=blue,
    urlcolor=magenta,
} 
\newcommand{\dd}{{\rm d}}
\newcommand{\be}{\begin{equation}}
\newcommand{\ee}{\end{equation}}
\newcommand{\bea}{\begin{eqnarray}}
\newcommand{\eea}{\end{eqnarray}}

\newcommand{\geff}{G_{\rm eff}} 
\newcommand{\alb}{\alpha_B} 
\newcommand{\alm}{\alpha_M} 
\newcommand{\al}{\alpha} 
\newcommand{\ms}{M_\star^2} 
\newcommand{\lcdm}{$\Lambda$CDM} 
\newcommand{\fsig}{f\sigma_8} 
\newcommand{\cltg}{C_{\ell}^{Tg}} 

\begin{document}

\title{Modified Gravity Away from a $\Lambda$CDM Background}

\author{Guilherme Brando${}^{*1}$, Felipe T.\ Falciano${}^{1,2}$, Eric V.\ Linder${}^{3,4}$, Hermano E.\ S.\ Velten${}^{1,5}$}
\affiliation{
${}^1$PPGCosmo, CCE - Universidade Federal do Esp\'irito Santo,
zip 29075-910, Vit\'oria, ES, Brazil\\
${}^2$CBPF - Centro Brasileiro de Pesquisas F\'isicas, Xavier Sigaud st.\ 150, zip 22290-180, 
Rio de Janeiro, RJ, Brazil\\ 
${}^3$Berkeley Center for Cosmological Physics \& Berkeley Lab, 
University of California, Berkeley, CA 94720, USA\\ 
${}^4$Energetic Cosmos Laboratory, Nazarbayev University, 
Nur-Sultan, Kazakhstan 010000\\ 
${}^5$Departamento de F\'isica, Universidade Federal de Ouro Preto (UFOP), zip 35400-000, Ouro Preto, MG, Brazil
}
\email{gbrando@cosmo-ufes.org}

\date{\today}

\begin{abstract}
Within the effective field theory approach to cosmic acceleration, the background 
expansion can be specified separately from the gravitational modifications. We 
explore the impact of modified gravity in a background different from a cosmological constant plus 
cold dark matter ($\Lambda$CDM) on the stability and cosmological observables, 
including covariance between gravity and expansion parameters. In No Slip 
Gravity the more general background allows more gravitational freedom, including 
both positive and negative Planck mass running. We examine the effects on cosmic structure 
growth, as well as showing that a viable positive integrated Sachs-Wolfe effect 
crosscorrelation easily arises from this modified gravity theory. Using current 
data we constrain parameters with a Monte Carlo analysis, finding a  
bound on the running $|\alpha_{M,{\rm max}}|\lesssim 0.03$ (95\% CL) for the adopted form at all cosmic times.
We provide the modified {\tt hi\_class} code publicly 
on GitHub, now enabling computation and inclusion of the 
redshift space distortion observable $\fsig$ as well as the 
No Slip Gravity modifications. 
\end{abstract} 

\maketitle

%%%%%%%%%%%%%%%%%%%%%%%%%%%%%%%%%%%%%%%%%%%%%%%%%%%%%%%%%%%% 
\section{Introduction}

Cosmic acceleration arises from an unknown physical origin but leaves 
concrete signatures in cosmic distances, growth of structure, light propagation 
and lensing, and cosmic microwave background (CMB) anisotropies. Careful 
investigation of all of these can provide insight into whether the effects are 
wholly due to a change in the cosmic expansion rate or also modification of the 
strength of gravity. 

The background expansion in modified gravity theories, however, tends to be chosen as 
that of a cosmological constant plus cold dark matter (\lcdm), or solved for 
only in the simplest viable models, such as $f(R)$, where it lies very close to 
\lcdm. However, the expansion rate is a function to be specified in the theory, 
just as the perturbative effective field theory or property functions are 
\cite{gubitosi2014,bloomfield2014,gleyzes2014,bellinisawicki,linder2016}. One can also choose to work from a given Lagrangian and compute expansion and perturbations together, though one has then to check that the expansion describes the data. We follow the common path of specifying the expansion separately to ensure it is viable.  Here 
we examine the implications of allowing background cosmologies away from \lcdm, 
as well as modified gravity, and their interplay. 

Of particular interest is how this affects cosmic growth observables, which 
depend both on the expansion rate and strength of gravity, and the 
crosscorrelation of perturbed quantities, such as CMB temperature anisotropies 
from the integrated Sachs-Wolfe (ISW) effect and galaxy clustering density. 
Indeed, some theories have been ruled out due to possessing an anticorrelation for this, 
rather than the observed positive correlation. Theories can also be discarded 
ab initio if they are unstable, but a non-\lcdm\ background offers extra possibilities for 
stabilizing some theories. 

The range of allowed effective theories is large, even with the tensor sector 
constrained to have the speed of gravitational waves equal to the speed of light. 
Therefore we consider particular connections between the two relevant property 
functions -- the Planck mass running and the kinetic braiding. A specific 
instantiation of such a relation is No Slip Gravity \cite{nsg}, one of the 
simplest and most predictive modified gravity theories, and we use this as an 
exemplar for the detailed calculations. 

In Sec.~\ref{sec:stable} we briefly review the property function formalism and 
explore the space of stable theories, also considering viability in terms of 
CMB observations. Section~\ref{sec:nsg} examines more closely No Slip Gravity in 
a non-\lcdm\ background, showing how the parameter space is enlarged. We 
investigate the impact on the cosmic structure growth rate in Sec.~\ref{sec:fsig}, 
and the lensing potential and ISW effect in Sec.~\ref{sec:isw}. 
Section~\ref{sec:mcmc} presents a Markov Chain Monte Carlo analysis of current 
data and constrains background and gravity parameters simultaneously. We conclude 
in Sec.~\ref{sec:concl}.

%%%%%%%%%%%%%%%%%%%%%%%%%%%%%%%%%%%%%%%%%% 
\section{Gravity in a non-\lcdm\ Background} \label{sec:stable} 

A convenient formalism for exploring many theories of cosmic modified gravity was 
developed by \cite{bellinisawicki}, involving four property functions, and the expansion 
history $H(a)$. These completely characterize the theory at the linear perturbation level. While this is an impressive simplification when working with Horndeski's most general 
scalar-tensor gravity theory \cite{horn,defa,bellinisawicki} or the effective field 
theory of dark energy \cite{gubitosi2014,bloomfield2014,gleyzes2014,linder2016}, this 
still leaves five free functions of time to specify. 

The detection of a binary neutron star merger with gravitational waves \cite{abbott2017i} and its 
electromagnetic counterparts \cite{abbott2017ii,abbott2017iii} provided a constraint on 
the speed of propagation of gravitational waves $c_{T}^{2}=1+\alpha_{T}$, with 
$\alpha_{T}=0$ in the most straightforward interpretation. Another property function, 
the kineticity $\al_K$, has little effect on subhorizon physics and generally does not 
need to be specified in detail. This leaves the Planck mass running $\al_M$ and the 
braiding $\al_B$, as well as the background itself, e.g.\ the Hubble parameter $H(a)$, 
where $a$ is the cosmic expansion factor. 

The arbitrariness and generality of the functional form of the $\alpha_{i}(a)$ 
functions can lead the theory to unphysical regimes. Three types of instabilities can 
violate the soundness of the theory: tachyon, ghost, and gradient. As pointed out, and  carefully analyzed in \cite{frusciante2018}, the first type of instability is less 
pathological and is associated with the large scale, low-$k$ regime (where $k$ is the 
Fourier mode), and is commonly not directly used in the modified gravity Boltzmann codes 
available in the literature, such as EFTCAMB \cite{hu2014i,hu2014ii} and 
{\tt hi\_class}\ \cite{miguel2017i}. The other two instabilities are more severe, and 
must be avoided. This provides constraints on the $\alpha_{i}$ functions. For the no 
ghost condition, $\al_K+(3/2)\alb^2\ge0$, this is readily satisfied by choosing 
$\al_K>0$. 

Avoidance of gradient instabilities corresponds to the scalar sound speed squared being 
nonnegative, 
\begin{equation}\label{cs2gen}
\begin{split}
    c_{s}^{2} =& \frac{1}{\alpha_{K}+3\alpha_{B}^{2}/2} \left[\left(1-\frac{\alpha_B}{2}\right)\left(2\alpha_{M}+\alpha_{B}-2\frac{H'}{H}\right)\right.\\
    &\left.+\alpha'_B-\frac{\tilde{\rho}_{m}+\tilde{p}_{m}}{H^2}\right] \ge0\ ,
\end{split}
\end{equation}
where a prime is a derivative with respect to $\ln a$ and a tilde denotes division by 
$\ms(a)/M_{\rm Pl}^2$, where $\ms$ is the running Planck mass squared. In terms of an 
effective dark energy we can write 
\begin{equation}\label{cs2w0wa}
\begin{split} 
    c_{s}^{2} =& \frac{1}{\alpha_{K}+3\alpha_{B}^{2}/2} \left[\left(1-\frac{\alpha_B}{2}\right)\left(2\alpha_{M}+\alpha_{B}\right)\right.\\
    &\left.+\frac{(H\alpha_B)'}{H}+\frac{\rho_m}{H^2}\left(1-\frac{M_{\rm Pl}^2}{\ms}\right)+\frac{{\rho}_{\rm de}(1+w)}{H^2}\right] \ge0\ ,
\end{split}
\end{equation} 
where $w$ is the effective dark energy equation of state parameter. For a \lcdm\ 
background, $1+w=0$. 

Thus a change in the background changes the stability condition. 
Taking the  
example of No Slip Gravity, where $\alb=-2\alm$, the stability region \textbf{is} 
\be 
\frac{(\alm H)'}{H}\le \frac{3}{2}\,\Omega_{\rm de}(a)\,[1+w(a)]+\frac{3}{2}\,\left(\Omega_m(a)-\tilde\Omega_m(a)\right)\ , \label{eq:stabnoslip} 
\ee 
where $\tilde\Omega_{m}=\tilde\rho_{m}/(3H^2)$. In particular, while a \lcdm\ 
background requires $\alm\ge0$ for stability if gravity is strengthened ($M_{\rm Pl}^2/\ms>1$)
since $H'<0$ at all times in a normal cosmic history, in the enlarged space $\alm<0$ is also 
allowed. 

This provides a motivation for studying non-\lcdm\ backgrounds, since the enlarged 
parameter space may also lead to different observational characteristics. For general 
time dependencies $\alm(a)$, $\alb(a)$, and $w(a)$ there is little specific that can 
be said, so we will have to parametrize these functions. For the effective dark energy 
we adopt the common $w(a)=w_0+w_a(1-a)$, which has been demonstrated to work for a 
broad class of scalar field and modified gravity theories. For $\alb(a)$ we explore 
the class of theories where this is proportional to $\alm(a)$, i.e.\ 
\be 
\alb(a)=-r\,\alm(a)\ . \label{eq:propr} 
\ee 
Such a relation holds for No Slip Gravity ($r=2$) and $f(R)$ gravity, Brans-Dicke, and 
chameleon theories ($r=1$). 
The $\Lambda$CDM background case was studied in \cite{denissenya2018}. 

\begin{figure}[htbp!]
\includegraphics[width=\columnwidth]{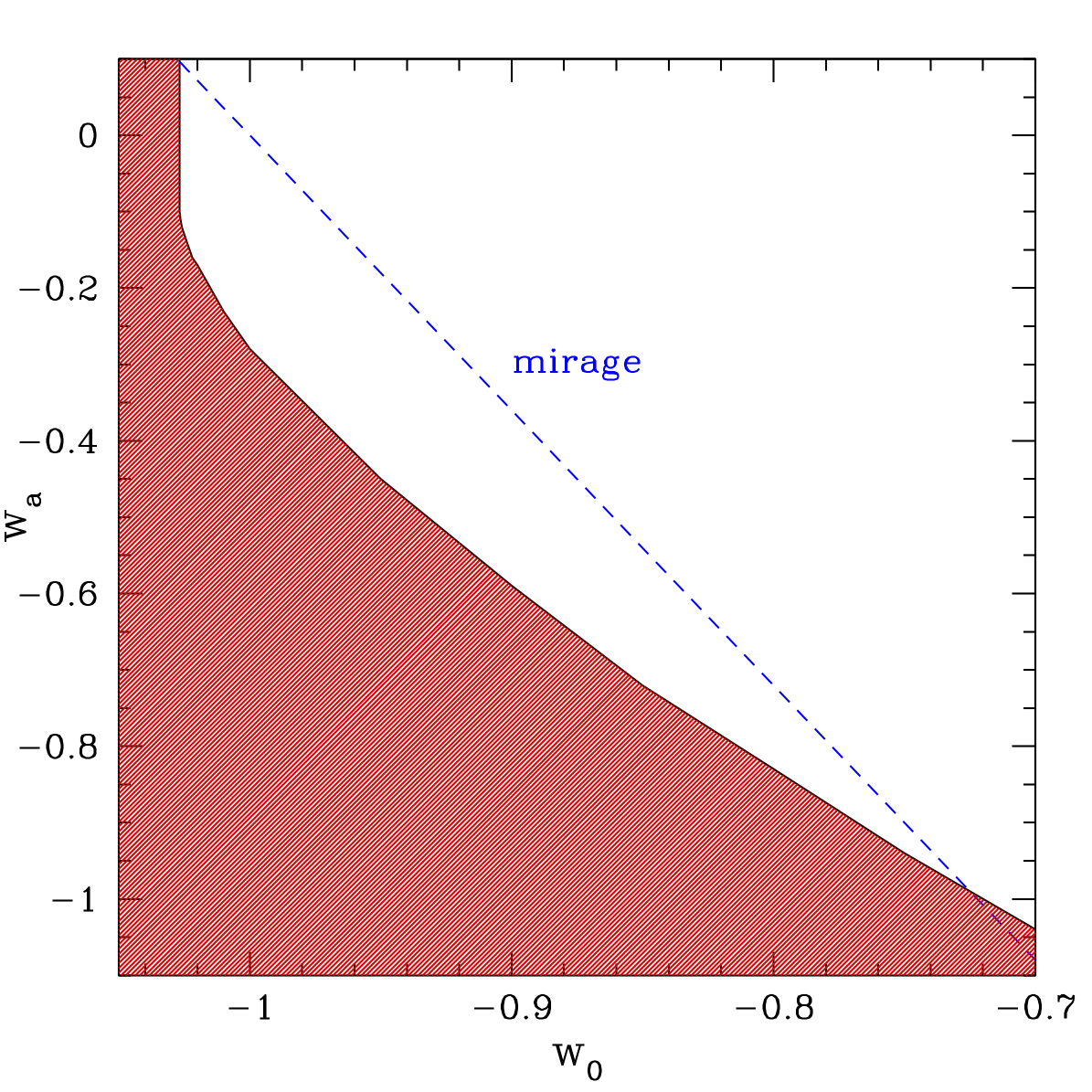} 
\caption{Stability region in the $w_0$--$w_a$ plane for No Slip Gravity with the 
hill/valley form of 
$\alm(a)$ with parameters $c_{M}=-0.05$, $\tau=1$, and $a_{t}=0.5$. Red regions indicate 
instability. The mirage relation $w_a=-3.6(1+w_0)$ is plotted as the dashed blue line.}
\label{fig:stabnoslipw0wa}
\end{figure}

Figure~\ref{fig:stabnoslipw0wa} shows the stability region for $a\le1$ in the 
$w_0$--$w_a$ parameter space for the example of No Slip Gravity. The \lcdm\ value 
$(w_0,w_a)=(-1,0)$ is stable and a significant part of the region $w_0>-1$ is as well. 
There is a sharp boundary 
as $w_0$ gets appreciably smaller than $-1$ (roughly $w_0<-1.026$ for the $\alm$ 
parameters used; this is independent of $w_a$ because the instability arises at late 
times, i.e.\ $a=1$). The form of $\alm(a)$ used here is the hill/valley form discussed 
below (a similar picture holds for the hill form of \cite{nsg}, also discussed below). 
We also indicate the mirage relation $w_a=-3.6\,(1+w_0)$ that nearly preserves the 
\lcdm\ distance to CMB last scattering \cite{mirage} and so indicates a level of 
observational viability.

%%%%%%%%%%%%%%%%%%%%%%%% 

\begin{figure}[htbp!]
\includegraphics[width=\columnwidth]{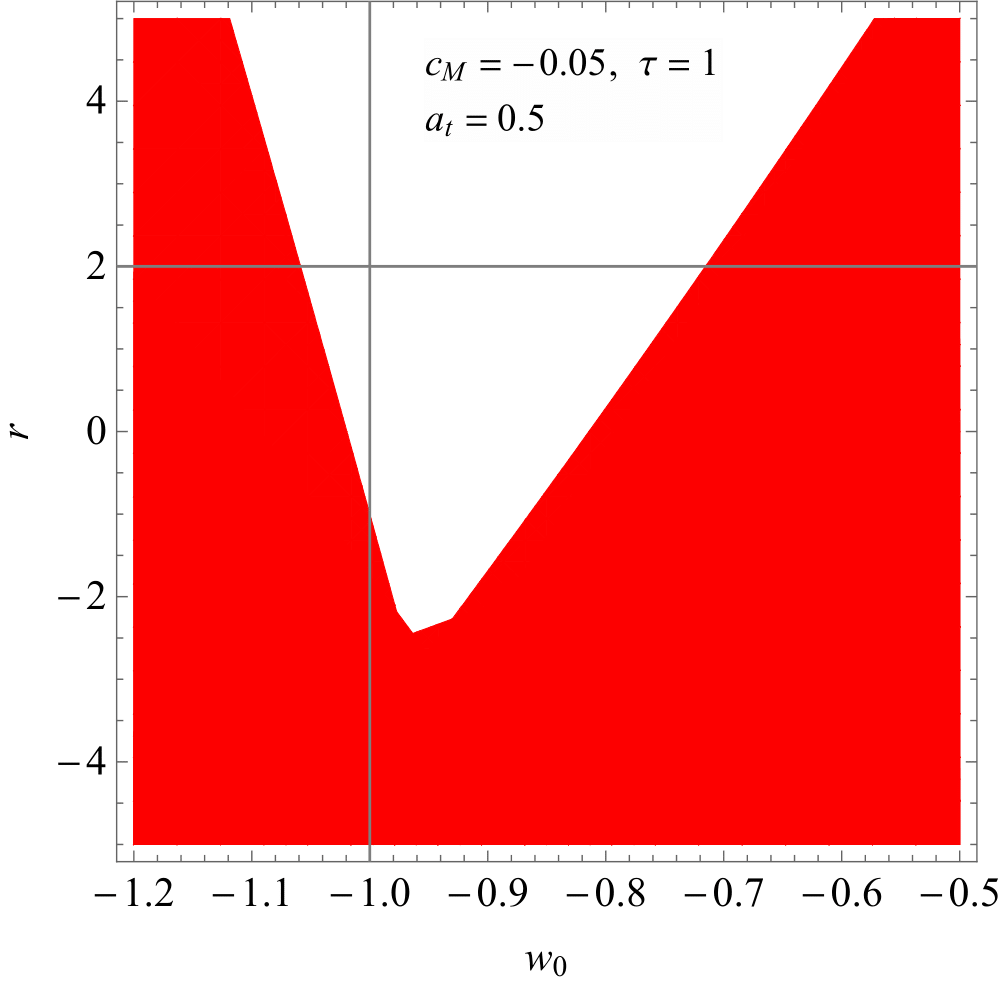} 
\caption{Stability region in the $r$--$w_0$ plane for $w_{a}$ given by the mirage 
relation. Red regions give instability. This adopts the hill/valley form for $\alm(a)$ 
with parameters $c_{M}=-0.05$, $\tau=1$ and $a_{t}=0.5$. The crosshairs center on  
No Slip gravity in a \lcdm\ background. }  
\label{fig:stabrmirage}
\end{figure}

Alternatively, Fig.~\ref{fig:stabrmirage} shows the stability region as we allow $r$ to 
vary, but restrict the dark energy equation of state to the mirage form. (Allowing $r$, 
$w_0$, and $w_a$ all to be free adds little qualitatively and diminishes the clarity of 
the plots.) As $r$ gets large the stable parameter space opens up in $w_0$--$w_a$ 
(for this hill/valley form of $\alm(a)$ at least). Note that $r\to\infty$, i.e.\ 
$\alm=0$ but $\alb\ne0$, corresponds to No Run Gravity \cite{nrg}.

%%%%%%%%%%%%%%%%%%%%%%%%%%% 

%%%%%%%%%%%%%%%%%%%%%%%%%%%%%%%%%%%%%%%%%%%% 
\section{No Slip Gravity} \label{sec:nsg} 

For the remainder of the article we focus on No Slip Gravity, as an intriguingly minimal 
modification with interesting phenomenology (e.g.\ suppression of growth, unusual for 
modified gravity) and good stability. Note that even with a change in background, the 
no slip condition remains $\alb=-2\alm$. 
%As mentioned, for stability in a \lcdm\ background 
%the $\alm$ property function must satisfy \cite{nsg} 
%\begin{equation}
%    \frac{(\alpha_{M} H)'}{H} = \alm'+\alpha_{M}\frac{H'}{H}\le 0\ , 
%\end{equation}
%which implies
%\begin{equation}
%    \alpha_{M}\ge 0\ ,
%\end{equation}
%since $H'<0$ at all times in a normal cosmic history. However, 

Since Eq.~(\ref{eq:stabnoslip}) 
allows $\alm<0$ as the right hand side can be lifted off zero, this opens a window for 
negative $\alm$ at some point in its evolution. 

We therefore change the hill form of \cite{nsg} where 
\bea 
\alm(a)&=&c_M\left(1-\tanh^2\left[\frac{\tau}{2}\ln\frac{a}{a_t}\right]\right)\notag\\ 
&=&\frac{c_M}{\cosh^2[(\tau/2)\ln(a/a_t)]}\notag\\ 
&=&\frac{4c_M\,(a/a_t)^\tau}{[(a/a_t)^\tau+1]^2} \ , \label{eq:hill} 
\eea  
to allow for a negative part of $\alm(a)$, i.e.\ a valley as well as a hill. 
That is, the theory changes qualitatively to permit both positive and 
negative Planck mass running during the evolution. The simplest modification incorporating this 
change without adding any further parameters we call the 
hill/valley form: 
\begin{equation} 
\begin{split}
   \alpha_{M}(a) &= c_{M}\frac{\tanh\left[ (\tau/2) \ln(a/a_{t}) \right]}{\cosh^{2}\left[ (\tau/2) \ln(a/a_{t}) \right]} \\
   &= \frac{4c_{M} (a/a_{t})^{\tau} \left[ -1 + (a/a_{t})^{\tau} \right]}{\left[1+(a/a_{t})^{\tau}\right]^{3}}\ .\label{eq:hillv} 
\end{split}
\end{equation} 
This illustrative form has the key characteristic of both positive and negative $\alm$ during evolution, 
while retaining the flexibility to adjust the amplitude (through $c_M$), the breadth of the 
behavior (through $\tau$), and the time of the transition (through $a_t$).

In the early universe $\alpha_{M} \approx -4c_{M} (a/a_{t})^{\tau}$, so we want $\tau>0$ 
to preserve general relativity at early times. 
(Formally one can switch the signs of $\tau$ and $c_M$, as seen in the first equation 
above, and get the same results; we take the $\tau>0$ branch.) 
The function then dips into a valley / 
rises to a hill for $c_M>0$ / $c_M<0$. At late times, in the far future $a\gg a_t$, the 
running vanishes as $(a/a_t)^{-\tau}$. This is as expected for a de Sitter asymptote 
but not required for $w\ne-1$ backgrounds. However, we only apply this form to past 
history, $a\le1$, where there are observational constraints. The parameters are 
$c_M$, related to the amplitude, $a_t$ is the scale factor of the transition 
between valley and hill (with $\alm(a_t)=0$), and $\tau$ measures the rapidity of the 
transition. Note that unlike the hill form, $c_M$ is not the maximum amplitude; rather, 
the extreme (maximum and minimum) amplitudes are 
\be \label{alMmax}
\al_{M,{\rm ext}}=c_M\,\frac{10\pm6\sqrt{3}}{27\pm15\sqrt{3}}\approx \pm 0.385\,c_M\ . 
\ee 
The depth of the valley and height of the hill agree, and these occur symmetrically 
around $a_t$, with 
\be 
a_{\rm max}=a_t\,(2+\sqrt{3})^{1/\tau}=\frac{a_t^2}{a_{\rm min}}\ . 
\ee 
For $\tau=1$ we have $a_{\rm max}=3.73a_t$, $a_{\rm min}=0.27a_t$. 

From $\alm(a)$ one derives the Planck mass squared $\ms$ through 
\be 
\frac{\ms}{M_{\rm Pl}^2}=e^{\int_0^a da'/a'\,\alm(a')}\ . 
\ee 
For the hill/valley form this becomes 
\begin{equation}\label{eq:ms}
\frac{M^{2}_{\star}}{M_{\rm Pl}^{2}} = \exp \left[\frac{-4(c_{M}/\tau)(a/a_{t})^{\tau}}{\left[1+ (a/a_{t})^{\tau}\right]^{2} }\right] \ . 
\end{equation} 
This smoothly evolves from 1 in the early universe to an extremum at $a=a_t$ with 
$\ms(a_t)/M_{\rm Pl}^2=e^{-c_M/\tau}$ and then back to 1 in the far future. 

Note that in No Slip Gravity the modified gravitational strengths in 
the matter and relativistic particle (light) Poisson equations are 
\be 
\geff\equiv G_{\rm matter}=G_{\rm light}=\frac{M_{\rm Pl}^2}{\ms}\ . 
\ee 
Whether $\ms$ grows initially (weaker gravity) or 
diminishes (stronger gravity) depends on the sign of $c_M$. Stability requires $\alm>0$ 
in the early universe and so we must have $c_M<0$. Thus the interesting feature of weaker 
gravitational strength from No Slip Gravity holds even in a non-\lcdm\ background. 

\begin{figure}[htbp!]
\includegraphics[width=\columnwidth]{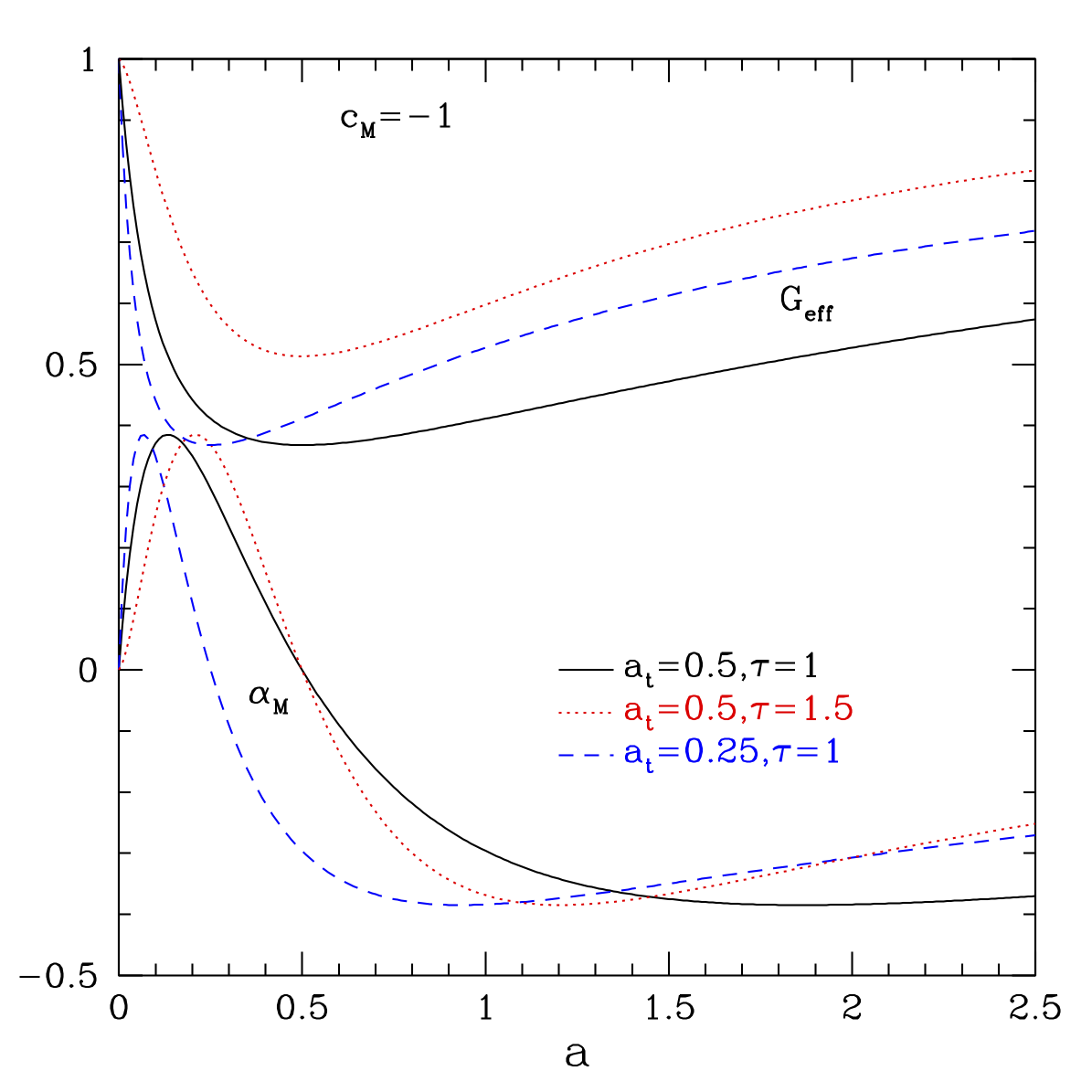} 
\caption{Curves of $\geff$ and $\alpha_{M}$ for the hill/valley form are shown for different values of $\tau$ and $a_t$, with $c_M=-1$. Positive $c_M$ 
reflects $\alm$ about 0, so hills become valleys, 
and inverts $\geff$, so values less than one 
become greater than one. 
}
\label{fig:almgeffa}
\end{figure}

Figure~\ref{fig:almgeffa} shows $\alm(a)$ and $\geff(a)$ for different values 
of the hill/valley parameters. Changing $a_t$ affects when $\alm$ crosses zero, 
i.e.\ the transition time between the hill and valley. Increasing $\tau$ steepens 
the transition, moving the minimum and maximum values of $\alm$ closer to the 
zero crossing. The amplitude of $\alm$ is governed by $c_M$, scaling linearly 
with it. Inverting the sign of $c_M$ would change hills to valleys and vice versa. 
For $\geff$, we see that indeed for $c_M<0$ gravity is weakened, where unity 
corresponds to the gravitational strength being Newton's constant. The maximum weakening occurs at 
$a_t$. Since $\geff$ returns to unity for scale factors $a\gg a_t$, then smaller 
$a_t$ means $\geff$ deviates from general relativity for a shorter time. 
Increasing $\tau$ again squeezes the transition, but also affects the maximum 
amplitude. Recall from Eq.~(\ref{eq:ms}) that the maximum deviation is 
$G_{\rm eff,max}=e^{c_M/\tau}$. Increasing $c_M$ increases the amplitude, 
exponentially. 

For illustrative purposes, the plots in the next two 
sections will fix $a_t=0.5$ and $\tau=1$ -- values near 
the edge of the eventual 68\% confidence limit joint 
posterior -- to more clearly show the effects of the 
modified gravity on observables. 
When we carry out Monte Carlo constraint analysis in 
Sec.~\ref{sec:mcmc} we will show the impact of fixing 
$a_t$ and $\tau$ vs fitting for $\{c_M,a_t,\tau\}$ 
simultaneously.

%%%%%%%%%%%%%%%%%%%%%%%%%%%%%%%%%% 

%%%%%%%%%%%%%%%%%%%%%%%%%%%%%%%%%%%%%%%% 
\section{Effects on Cosmic Growth} \label{sec:fsig} 

Changes to the strength of gravity, $\geff$, will directly affect the growth 
of large scale structure in the universe. This can be measured through galaxy 
redshift surveys through redshift space distortions caused by the velocities 
due to gravitational clustering, in the form of the cosmological parameter 
combination $\fsig(a)$. Here $f$ is the logarithmic growth rate and $\sigma_8$ 
is the mass fluctuation amplitude. 

For various cosmological backgrounds, i.e.\ expansion histories described by 
matter plus dark energy with 
a mirage equation of state, we solve numerically the subhorizon linear density perturbation growth 
equation with various modified gravitational strengths $\geff$. The solutions 
for the redshift space distortion (RSD) parameter $\fsig(a)$ of the growth rate history 
are compared to the equivalent result for the same background but with general 
relativity, and to current observational data. 

\begin{figure}[htbp!]
\includegraphics[width=\columnwidth]{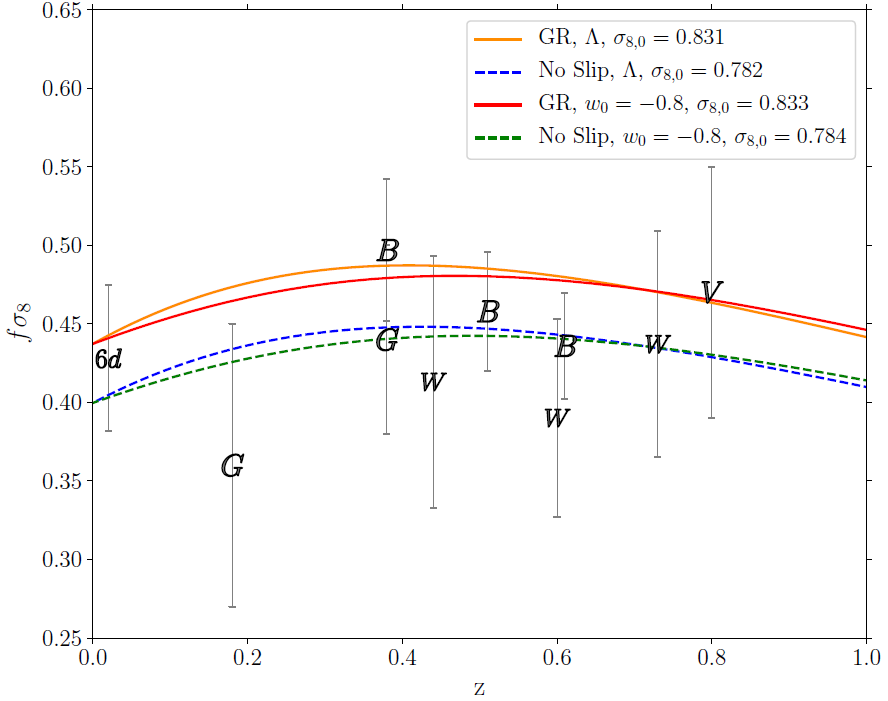} 
\caption{The redshift space distortion observable $f\sigma_{8}$, basically 
the growth rate history, is plotted for \lcdm\ and for mirage dark energy with 
present equation of state parameter $w_0$, in general relativity (GR) and in 
No Slip Gravity with $c_M=-0.05$, $a_t=0.5$, $\tau=1$. All curves have fixed 
$\Omega_{m,0}=0.314$ and the same initial conditions, and the derived values of $\sigma_{8,0}$ are indicated in the legend. Galaxy redshift 
survey data points are shown with their error bars. Note that No Slip Gravity 
suppresses growth, unlike many modified gravity theories, bringing the theory 
into better agreement with this growth data. 
} 
\label{fig:fsigw0}
\end{figure}

Figure~\ref{fig:fsigw0} shows the results. The observational data points come 
from the galaxy redshift surveys of 6dFGRS \cite{6dFGS}, GAMA \cite{GAMA}, BOSS 
\cite{baobossdr12ii}, WiggleZ \cite{WiggleZ}, and VIPERS \cite{VIPERS}. 
Indeed No Slip Gravity, even in the hill/valley form where 
$\alm$ can be both positive and negative during its evolution, 
suppresses growth relative to the general relativity with the 
same background expansion. This characteristic, rare for modified gravity theories, 
gives an improved fit to the RSD data for the same background. 
(To be absolutely proper, one should reanalyze the galaxy clustering data within 
the theory to be tested but this is beyond the scope of this paper and at the level of 
current data precision and small deviations from GR \lcdm\ this should not be a 
large effect.)

%%%%%%%%%%%%%%%%%%%%%%%%%%%%%%%%% 

We also see that the mirage 
dark energy models, even with an equation of state today as far 
from a cosmological constant as $w_0=-0.8$, have quite similar 
growth histories as in the corresponding \lcdm\ model of the same gravitational 
theory, i.e.\ general relativity or No Slip Gravity. This is one of the useful properties of the 
mirage models, even in the nonlinear power spectrum, as highlighted in 
\cite{0704.0312,mirage}.

%%%%%%%%%%%%%%%%%%%%%%%%%%%%%%%%%%%%%%%% 
\section{Lensing Potential and ISW Effect} \label{sec:isw} 

While we have considered the effect of modified gravity on the growth of cosmic 
structure, gravity also affects light propagation. That is, in addition to 
$G_{\rm matter}$ there is a modification of Poisson equation involving the sum of 
the metric potentials $\Phi+\Psi$ (often called the Weyl potential), or 
$G_{\rm light}$. Recall that for No Slip Gravity $G_{\rm light}=M^2_{\rm Pl}/\ms$. 
The sum of potentials generally decays in a universe with dark 
energy as matter domination wanes. However, if gravity is strengthened then it 
could overcome this tendency and grow the potentials. This not only gives a large 
integrated Sachs-Wolfe (ISW) effect (proportional to $\dot\Phi+\dot\Psi$) in the 
CMB but can cause an anticorrelation between the ISW and the density perturbations. 

Such issues are discussed in detail in \cite{renk2017i,renk2017ii,noller2018ii}, 
and some cubic Horndeski gravity theories indeed have a negative crosscorrelation 
between CMB temperature perturbations and galaxy density perturbations, 
$\cltg$. This conflicts with the prediction of \lcdm, and data, and is a strong 
indicator against such theories. (We note, however, that we have verified that 
No Run Gravity \cite{nrg}, a subclass of cubic Horndeski gravity, and with a 
strengthening of gravity, still does have a positive crosscorrelation.) 

\begin{figure}[htbp!]
\includegraphics[width=\columnwidth]{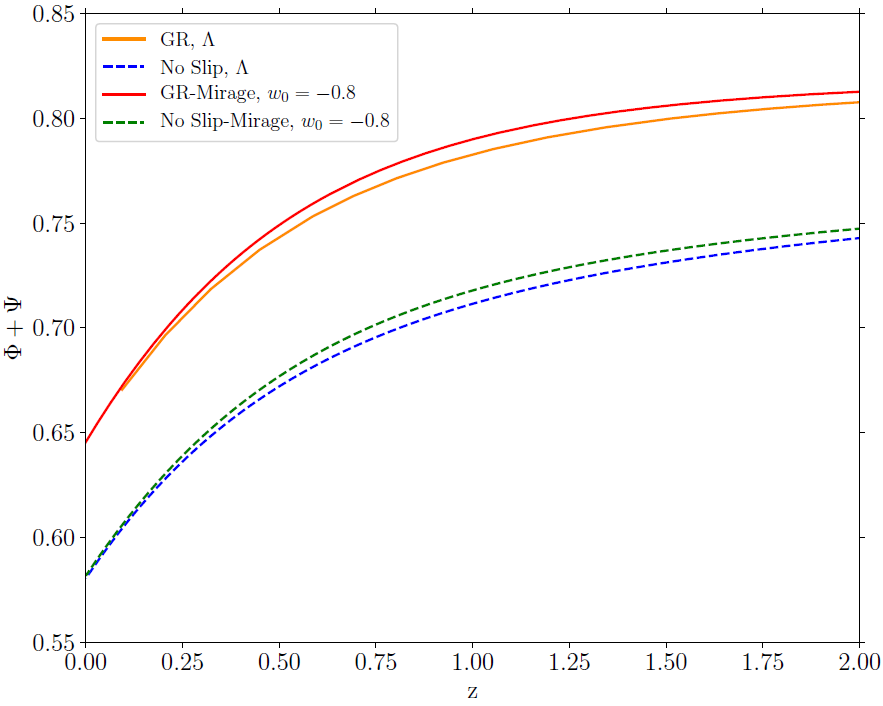}\\ 
\caption{The Weyl lensing potential, in CLASS code units, 
is plotted for \lcdm\ and for mirage dark energy with 
present equation of state parameter $w_0$, in general relativity (GR) and in 
No Slip Gravity with $c_M=-0.05$, $a_t=0.5$, $\tau=1$. The weakened gravity in 
No Slip Gravity enhances the decay of the potential, in contrast to, e.g., 
Galileon gravity. 
}
\label{fig:Weyl}
\end{figure}

Since No Slip Gravity weakens gravity, suppressing growth, we expect the Weyl 
potential to decay (i.e.\ weaker gravitational lensing). Figure~\ref{fig:Weyl} 
confirms this. The lensing potential in No Slip Gravity is suppressed relative 
to general relativity for the same background. 
(Note that at high redshift the curves approach the general relativity 
behavior.)  
One can use the same analytic 
calculation as in \cite{brush2018} to approximate the degree of suppression. Note 
that, as for growth, the mirage models act in light propagation quite similarly 
to the \lcdm\ model they were designed to mimic in CMB distance to last scattering.

%%%%%%%%%%%%%%%%%%%%%%%%%%%%%%%%%%%%%%% 

Given the preservation of the characteristic of a decaying lensing potential as 
in \lcdm, we might expect a positive temperature-density crosscorrelation at 
large angles (low multipoles $l$) where the ISW effect dominates. Let us 
calculate this in detail. 
We will follow closely the procedure outlined in \cite{renk2017ii}, to compute the cross correlation between the CMB temperature and a galaxy survey. First we must calculate
\begin{equation}
    C_{l}^{Tg} = 4 \pi \int \frac{\dd k}{k} \Delta_{l}^{ISW}(k) \Delta_{l}^{g}(k) \mathcal{P}_{\mathcal{R}}(k)\ ,
\end{equation}
where $\mathcal{P}_{\mathcal{R}}$ is the power spectrum of the primordial curvature perturbations ($\mathcal{R}(\mathbf{k})$), and $\Delta_{l}^{ISW}$ and $\Delta_{l}^{g}$ are the transfer functions for the ISW effect and for the galaxies. The first is given by
\begin{equation}
    \Delta_{l}^{\rm ISW} = \int_{\eta_{*}}^{\eta_{0}} \dd \eta\, (\Phi^{\prime}+\Psi^{\prime}) j_{l}\ , 
\end{equation}
where $\eta_{*}$ and $\eta_{0}$ are the conformal time at recombination and today, respectively, and a prime here denotes a derivative with respect to $\eta$. 
The transfer functions depend on the modified gravity theory being considered and 
are calculated through the perturbation equations, which are solved numerically by 
{\tt hi\_class}. 

For computations in which source number counts are present, the relevant transfer function 
is given as 
\begin{equation}
    \Delta_{l}^{g} \approx \Delta_{l}^{{\rm Den}_{i}} + \dots\ , 
\end{equation}
where the dots represent other contributions such as redshift-space distortions, lensing, polarization, and contributions suppressed by $H/k$ in subhorizon scales \cite{renk2017ii}. 
The explicit form of $\Delta_{l}^{{\rm Den}_{i}}$ is
\begin{equation}
    \Delta_{l}^{{\rm Den}_{i}} = \int_{0}^{\eta_{0}} \dd \eta\, W_{i} b_{g}(\eta) \delta(\eta,k) j_{l}\ ,
\end{equation}
where $\delta(\eta,k)$ is the density perturbation at the Fourier mode $k$, $j_{l}=j_{l}(k(\eta_{0}-\eta))$ is a Bessel function, and $W_{i}$ is a window function, discussed below. To be consistent with {\tt hi\_class} all transfer functions are normalized to the value of the curvature perturbation at some time $k\eta_{\rm ini} \ll 1$, e.g.\ $\delta(\eta, k)= \delta(\eta, \mathbf{k})/\mathcal{R}(\eta_{\rm ini},\mathbf{k})$.\\

For a galaxy sample we use the NVSS survey \cite{NVSS}, which covers the  
sky north of −40 deg declination in one band. This is a large area, fairly deep 
survey with good overlap with the CMB ISW kernel. The selection function $W_{i}$ is 
given by the observed number of sources per redshift, $dN/dz$, and we use a 
constant bias factor for each redshift bin. The survey selection function is given 
by \cite{hirata2008} as 
\begin{equation}\label{dNdzNVSS}
    \left[ b_{g}(z) \frac{\dd N}{\dd z} \right]_{\rm NVSS} = b_{\rm eff} \frac{\alpha^{\alpha+1}}{z_{0}^{\alpha+1}\Gamma(\alpha)}z^{\alpha} e^{-\alpha z/z_{0}}\ ,
\end{equation}
with $b_{\rm eff} = 1.98$, $z_{0} = 0.79$, $\alpha = 1.18$, and $\Gamma$ the gamma function.

\begin{figure}[htbp!]
\includegraphics[width=\columnwidth]{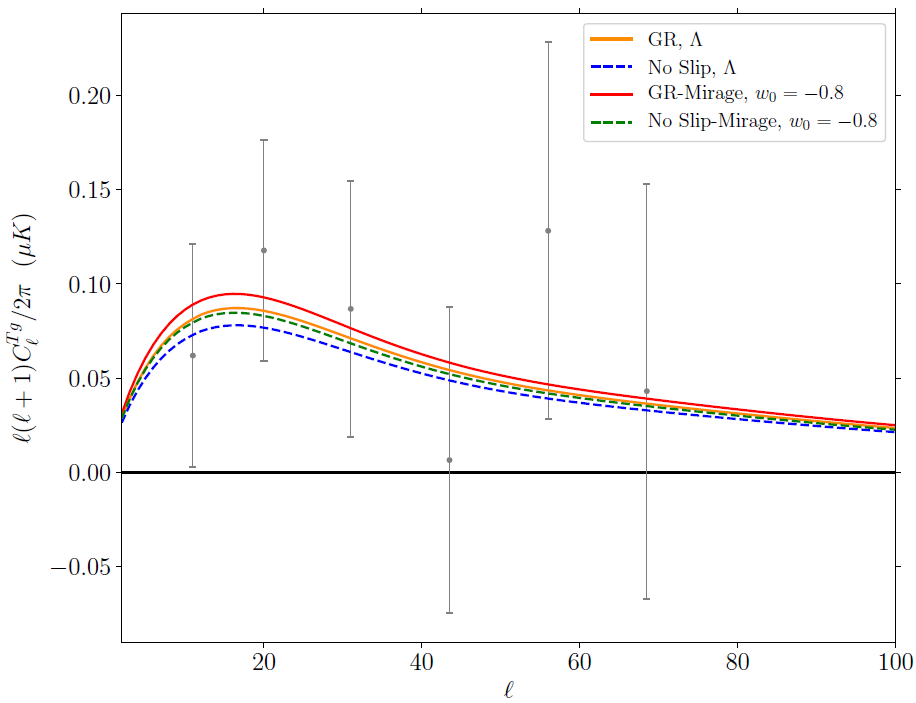}
\caption{The ISW-galaxy crosscorrelation $\cltg$ 
is plotted for \lcdm\ and for mirage dark energy with 
present equation of state parameter $w_0$, in general relativity (GR) and in 
No Slip Gravity with $c_M=-0.05$, $a_t=0.5$, $\tau=1$. 
The data points come from the NVSS survey, as extracted from \cite{hirata2008}. 
We see that, indeed, No Slip Gravity gives a positive crosscorrelation. 
}
\label{fig:isw}
\end{figure}

We modified {\tt hi\_class} in order to implement (\ref{dNdzNVSS}) in a specific 
subroutine of the {\tt transfer} module. Figure~\ref{fig:isw} shows the results. 
We see that indeed No Slip Gravity gives a positive ISW crosscorrelation, in 
agreement with the $\Lambda$CDM case, and observational data. However, without a 
proper calibration of the bias factor for the NVSS survey in No Slip Gravity with 
this background, as done in \cite{renk2017ii} for the Galileon model, we cannot 
investigate in quantitative detail a likelihood analysis of the ISW data. This is 
left for future work. The calibration of the bias would affect the height and 
position of the hill present for $\ell<20$. Note that on those large scales there 
is also an influence of the value chosen for the $\al_K$ parameter. We have 
investigated this and find that for $\al_K=0.1$ the effect is less than 0.2\% 
for $\ell>20$, rising to 0.5\% for the lowest $\ell$ (relative to the 
corresponding case with $\al_K=10^{-4}$). Given the size of the uncertainties in 
the data (including cosmic variance), this is a negligible effect.

%%%%%%%%%%%%%%%%%%%%%%%%%%%%%%%%%%%%%%% 

%%%%%%%%%%%%%%%%%%%%%%%%%%%%%%%%%%%%%%%%%%%%
\section{Cosmology and Gravity Constraints} \label{sec:mcmc} 

Having explored the impact of modified gravity in a non-\lcdm\ 
background on both growth of structure and 
light propagation we now proceed to perform a Markov Chain Monte Carlo (MCMC) analysis 
of our model using MontePython \cite{montepythoni,montepythonii}. 
We fit over the standard cosmological parameters plus some additional effective dark 
energy and modified gravity ones: $w_0$ and $w_a$ for the background and $c_M$, $a_t$, and $\tau$ for 
modified gravity. We do not apply the mirage relation between $w_0$ and $w_a$, but 
we will find that it gives a reasonable fit to the MCMC joint confidence contour 
(also see Fig.~2 of \cite{1704.00762}). 
In one case we fix $a_t=0.5$, $\tau=1$ as fiducial values, for reasons 
given in Sec.~\ref{sec:nsg}, but we also allow them to vary in another case. 
The sum of the masses of the neutrinos (one massive and two massless) is fixed to 
0.06 eV. On the extra parameters we use flat priors of $w_0\in[-1.2,0]$, 
$w_a\in[-1,0.5]$, and $c_M\in[-0.1,0]$. When varying the modified gravity transition 
parameters we use $a_t\in[0.1,1]$ and $\tau\in[0.33,2.19]$ from stability and 
observational considerations. These priors are informed by 
the stability analysis in Appendix~\ref{sec:apxtau}. 

For data sets we use CMB (Planck $TTTEEE$ \cite{plancklowhighl} and lensing 
\cite{plancklensing}), BAO (BOSS DR12 \cite{baobossdr12ii}, SDSS DR7 MGS \cite{sdss}, 
6dFGS \cite{6dFGS}), RSD (BOSS DR12 \cite{baobossdr12ii}), and supernovae (JLA \cite{JLAfull}).  
Note that we added to {\tt hi\_class} the capability to compute the redshift space distortion observable $\fsig$, which it previously 
lacked, and included this in MCMC likelihood evaluation for No Slip modified gravity. The modified code is publicly accessible at 
\url{https://github.com/gbrandool/hi_class_public}. 

\begin{figure*}
\includegraphics[width=0.95\textwidth]{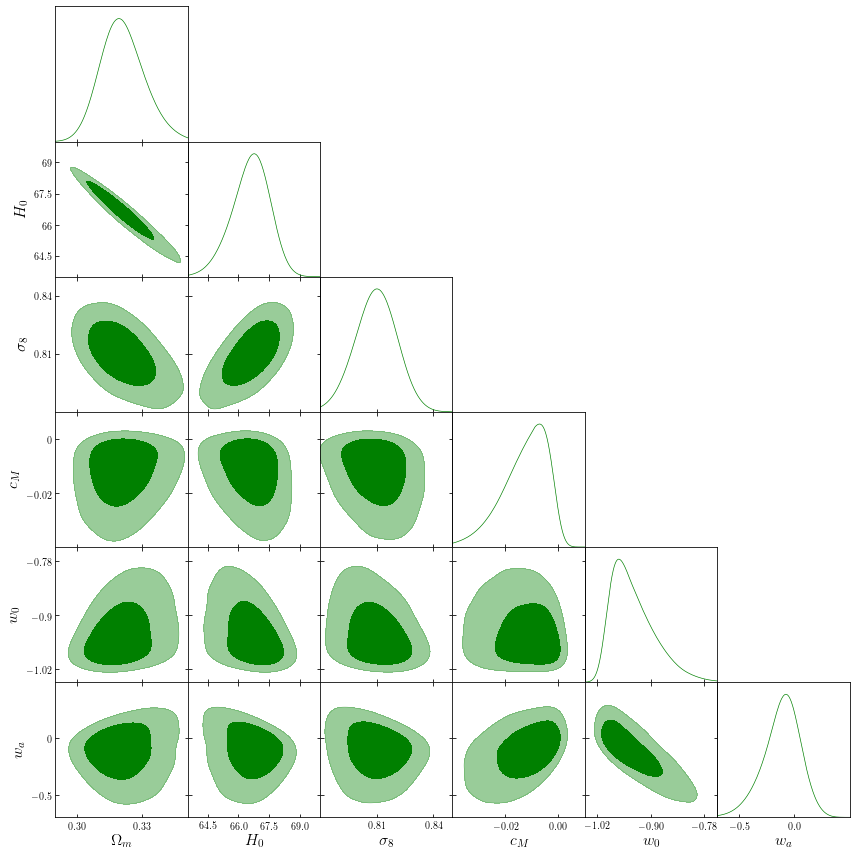}
\caption{Triangle plot of the joint probability distributions, at 68.3\% and 95.4\% confidence levels, and marginalized 
one dimensional posteriors, for various cosmological and gravity parameters. Here $a_{t}=0.5$ and $\tau=1$ are fixed. 
}
\label{fig:mcmcfixed}
\end{figure*}

All the parameter constraints were extracted using the Gelman-Rubin convergence 
diagnostic $R$, with a convergence criterion of $R-1<0.01$ \cite{gelmanrubin}. 
The derived constraints for the fixed $a_t$ and $\tau$ case are given in Table~\ref{tab:mcmcparamfixed} and the triangle plot 
in Figure~\ref{fig:mcmcfixed}.

%%%%%%%%%%%%%%%%%%%%%%%%%%%%%%%%%%%%%%%%%%%%%%
\begin{table}[htbp!]
\begin{tabular}{|l|c|c|c|c|}
 \hline
Param & best-fit & mean$\pm\sigma$ & 95\% lower & 95\% upper \\ \hline
\rule{0pt}{1.1\normalbaselineskip}$10^{2} \omega_b$ &$2.234$ & $2.225_{-0.016}^{+0.015}$ & $2.194$ & $2.257$ \\
\rule{0pt}{1.1\normalbaselineskip}$\omega_{\rm cdm}$ &$0.1193$ & $0.1194_{-0.0014}^{+0.0014}$ & $0.1167$ & $0.1222$ \\
\rule{0pt}{1.1\normalbaselineskip}$H_0$ &$67.75$ & $66.59_{-0.82}^{+1.1}$ & $64.63$ & $68.42$ \\
\rule{0pt}{1.1\normalbaselineskip}$10^{9} A_s$ &$2.163$ & $2.208_{-0.067}^{+0.058}$ & $2.085$ & $2.336$ \\
\rule{0pt}{1.1\normalbaselineskip}$n_s$ &$0.9647$ & $0.966_{-0.0046}^{+0.0046}$ & $0.9568$ & $0.9753$ \\
\rule{0pt}{1.1\normalbaselineskip}$\tau_{\rm reio}$ &$0.06993$ & $0.08_{-0.016}^{+0.014}$ & $0.05055$ & $0.1101$ \\
%\rule{0pt}{1.1\normalbaselineskip}$c_M$ &$-0.005683$ & $-0.01252_{-0.0032}^{+0.013}$ & $-0.02988$ & $-4.847\times 10^{-8}$ \\ 
\rule{0pt}{1.1\normalbaselineskip}$c_M$ &$-0.005683$ & $-0.01252_{-0.0032}^{+0.013}$ & $-0.02988$ & $0.0$ \\ 
\rule{0pt}{1.1\normalbaselineskip}$w_0$ &$-0.9953$ & $-0.9407_{-0.064}^{+0.023}$ & $-1.015$ & $-0.8363$ \\
\rule{0pt}{1.1\normalbaselineskip}$w_a$ &$-0.03216$ & $-0.1123_{-0.14}^{+0.19}$ & $-0.4807$ & $0.2184$ \\
\rule{0pt}{1.1\normalbaselineskip}$\sigma_8$ &$0.816$ & $0.8095_{-0.011}^{+0.012}$ & $0.7864$ & $0.8323$ \\
\hline
 \end{tabular}  \caption{Results of the MCMC analysis for various cosmological and gravity parameters, 
 for the case with $a_{t}=0.5$ and $\tau=1$ fixed. 
 } 
\label{tab:mcmcparamfixed}
\end{table}

\begin{table}[htbp!]
\begin{tabular}{|l|c|c|c|c|}
 \hline
Param & best-fit & mean$\pm\sigma$ & 95\% lower & 95\% upper \\ \hline
\rule{0pt}{1.1\normalbaselineskip}$10^{2} \omega_b$ &$2.227$ & $2.225_{-0.016}^{+0.016}$ & $2.193$ & $2.257$ \\
\rule{0pt}{1.1\normalbaselineskip}$\omega_{\rm cdm}$ &$0.119$ & $0.1195_{-0.0014}^{+0.0014}$ & $0.1167$ & $0.1224$ \\
\rule{0pt}{1.1\normalbaselineskip}$H_0$ &$67.38$ & $66.97_{-1.2}^{+1.1}$ & $64.62$ & $69.3$ \\
\rule{0pt}{1.1\normalbaselineskip}$10^{9} A_s$ &$2.156$ & $2.193_{-0.063}^{+0.056}$ & $2.073$ & $2.314$ \\
\rule{0pt}{1.1\normalbaselineskip}$n_s$ &$0.9662$ & $0.9656_{-0.0048}^{+0.0048}$ & $0.9561$ & $0.9749$ \\
\rule{0pt}{1.1\normalbaselineskip}$\tau_{\rm reio}$ &$0.06865$ & $0.0764_{-0.015}^{+0.014}$ & $0.0479$ & $0.106$ \\
%\rule{0pt}{1.1\normalbaselineskip}$c_M$ &$-0.004449$ & $-0.0322_{-0.009}^{+0.032}$ & $-0.07995$ & $-2.913\times 10^{-6}$ \\ 
\rule{0pt}{1.1\normalbaselineskip}$c_M$ &$-0.004449$ & $-0.0322_{-0.009}^{+0.032}$ & $-0.07995$ & $0.0$ \\ 
\rule{0pt}{1.1\normalbaselineskip}$a_t$ &$0.275$ & $0.676_{-0.094}^{+0.32}$ & $0.2615$ & $1.0$ \\
\rule{0pt}{1.1\normalbaselineskip}$\tau$ &$1.446$ & $1.631_{-0.15}^{+0.56}$ & $0.8304$ & $2.19$ \\
\rule{0pt}{1.1\normalbaselineskip}$w_0$ &$-0.9808$ & $-0.9358_{-0.073}^{+0.039}$ & $-1.04$ & $-0.7972$ \\
\rule{0pt}{1.1\normalbaselineskip}$w_a$ &$-0.04176$ & $-0.188_{-0.17}^{+0.29}$ & $-0.7417$ & $0.2638$ \\
\rule{0pt}{1.1\normalbaselineskip}$\sigma_8$ &$0.814$ & $0.8141_{-0.014}^{+0.013}$ & $0.7875$ & $0.8416$ \\
\hline
\end{tabular}
\caption{Results of the MCMC analysis for various cosmological and gravity parameters, 
 for the case with $a_{t}$ and $\tau$ varying. 
 } 
\label{tab:mcmcparamvary}
\end{table}

\begin{table}
\begin{tabular}{|l|c|c|c|c|}
 \hline
Param & best-fit & mean$\pm\sigma$ & 95\% lower & 95\% upper \\ \hline
\rule{0pt}{1.1\normalbaselineskip}$10^{-2} \omega_b$ &$2.229$ & $2.225_{-0.016}^{+0.015}$ & $2.194$ & $2.256$ \\
\rule{0pt}{1.1\normalbaselineskip}$\omega_{\rm cdm}$ &$0.1191$ & $0.1195_{-0.0014}^{+0.0013}$ & $0.1168$ & $0.1223$ \\
\rule{0pt}{1.1\normalbaselineskip}$H_0$ &$67.64$ & $67.43_{-0.58}^{+0.6}$ & $66.22$ & $68.61$ \\
\rule{0pt}{1.1\normalbaselineskip}$10^{9} A_s$ &$2.132$ & $2.177_{-0.064}^{+0.057}$ & $2.056$ & $2.302$ \\
\rule{0pt}{1.1\normalbaselineskip}$n_s$ &$0.9671$ & $0.9656_{-0.0046}^{+0.0045}$ & $0.9565$ & $0.9748$ \\
\rule{0pt}{1.1\normalbaselineskip}$\tau_{\rm reio}$ &$0.06472$ & $0.07305_{-0.015}^{+0.014}$ & $0.0436$ & $0.1031$ \\
\rule{0pt}{1.1\normalbaselineskip}$c_M$ &$-0.0002385$ & $-0.01762_{-0.0026}^{+0.018}$ & $-0.0556$ & $0.0$ \\
\rule{0pt}{1.1\normalbaselineskip}$a_t$ &$0.1929$ & $0.696_{-0.18}^{+0.24}$ & $0.31$ & $1.0$ \\
\rule{0pt}{1.1\normalbaselineskip}$\tau$ &$0.9227$ & $1.456_{-0.22}^{+0.73}$ & $0.6165$ & $2.19$ \\
\rule{0pt}{1.1\normalbaselineskip}$\sigma_8$ &$0.8146$ & $0.8176_{-0.009}^{+0.0092}$ & $0.7995$ & $0.836$ \\
\hline
 \end{tabular}
  \caption{Results of the MCMC analysis for various cosmological and gravity parameters, 
 for the $\Lambda$CDM case with $a_{t}$ and $\tau$ varying. 
 } 
\label{tab:mcmcparamvarylcdm}
\end{table}

The mass fluctuation amplitude $\sigma_8$ is lower than in the Planck analysis within general relativity, 
due to the suppression of growth by No Slip Gravity, as presaged in Fig.~\ref{fig:fsigw0}. 
This could put it in better agreement with weak lensing measurements 
\cite{sig8a,sig8b,sig8c,sig8d,eglow} (but see \cite{sig8hsc}), which 
are not included in this analysis. (Note the discussion in Sec.~\ref{sec:fsig} regarding 
formally needing to reanalyze data within the new theory.) The amplitude of the Planck mass running $\alm$, in 
terms of $c_M$, is restricted at the couple of percent level ($c_M>-0.03$ at 95\% CL), but this can still have a 
discernible effect on growth of structure and lensing. 
However general relativity ($c_M=0$) is within 
the 95\% confidence level. Again note the one sided distribution due to stability 
considerations. 

\begin{figure*}
\includegraphics[width=0.95\textwidth]{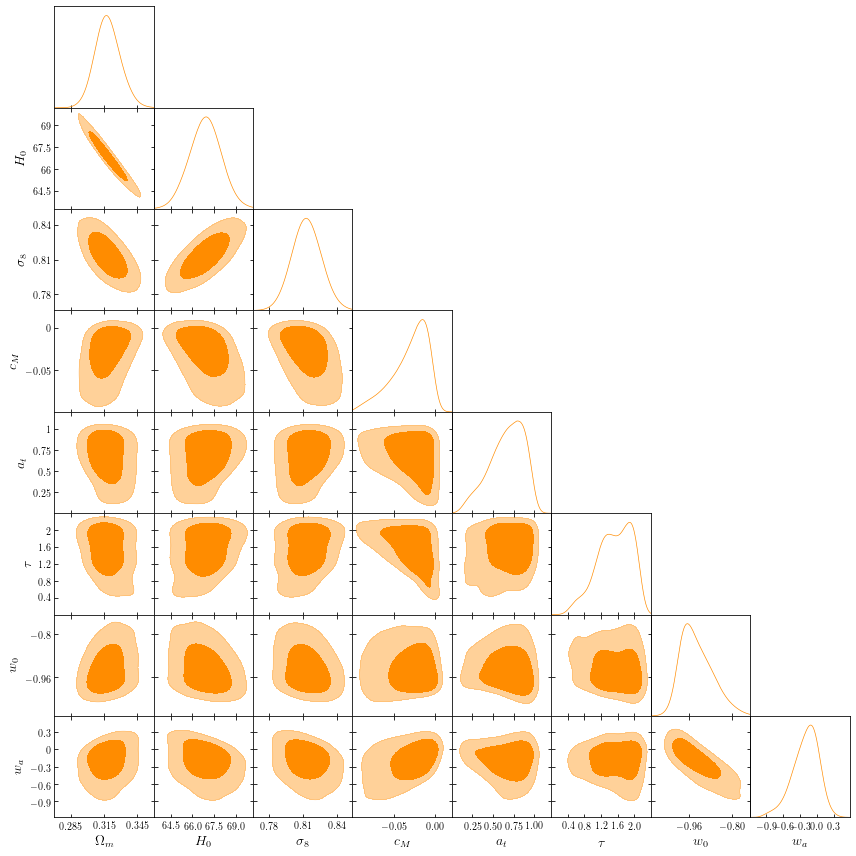}
\caption{Triangle plot of the joint probability distributions, at 68.3\% and 95.4\% confidence levels, and marginalized 
one dimensional posteriors, for various cosmological and gravity parameters. Here $a_{t}$ and $\tau$ are free to vary. 
}
\label{fig:mcmcvary}
\end{figure*}

We then repeat the analysis allowing $a_t$ and $\tau$ to vary.  The results are shown in Table~\ref{tab:mcmcparamvary} and in Fig.~\ref{fig:mcmcvary}. Note that the $a_t$ and $\tau$ posteriors have pulled away  
from the lower bounds on the priors (and the upper bounds are 
given by stability conditions). The exception is when $c_M$ 
approaches zero -- corresponding to general relativity -- where $a_t$ and $\tau$ become irrelevant, as seen from Eq.~(\ref{eq:hillv}). 
By allowing $a_{t}$ and $\tau$ to vary, $c_{M}$ can now assume more negative values than in the previous fixed case.  

%%%%%%%%%%%%%%%%%%%%%%%%%% 

%%%%%%%%%%%%%%%%%%%%%%%%%% 
\begin{figure*}
\includegraphics[width=0.95\textwidth]{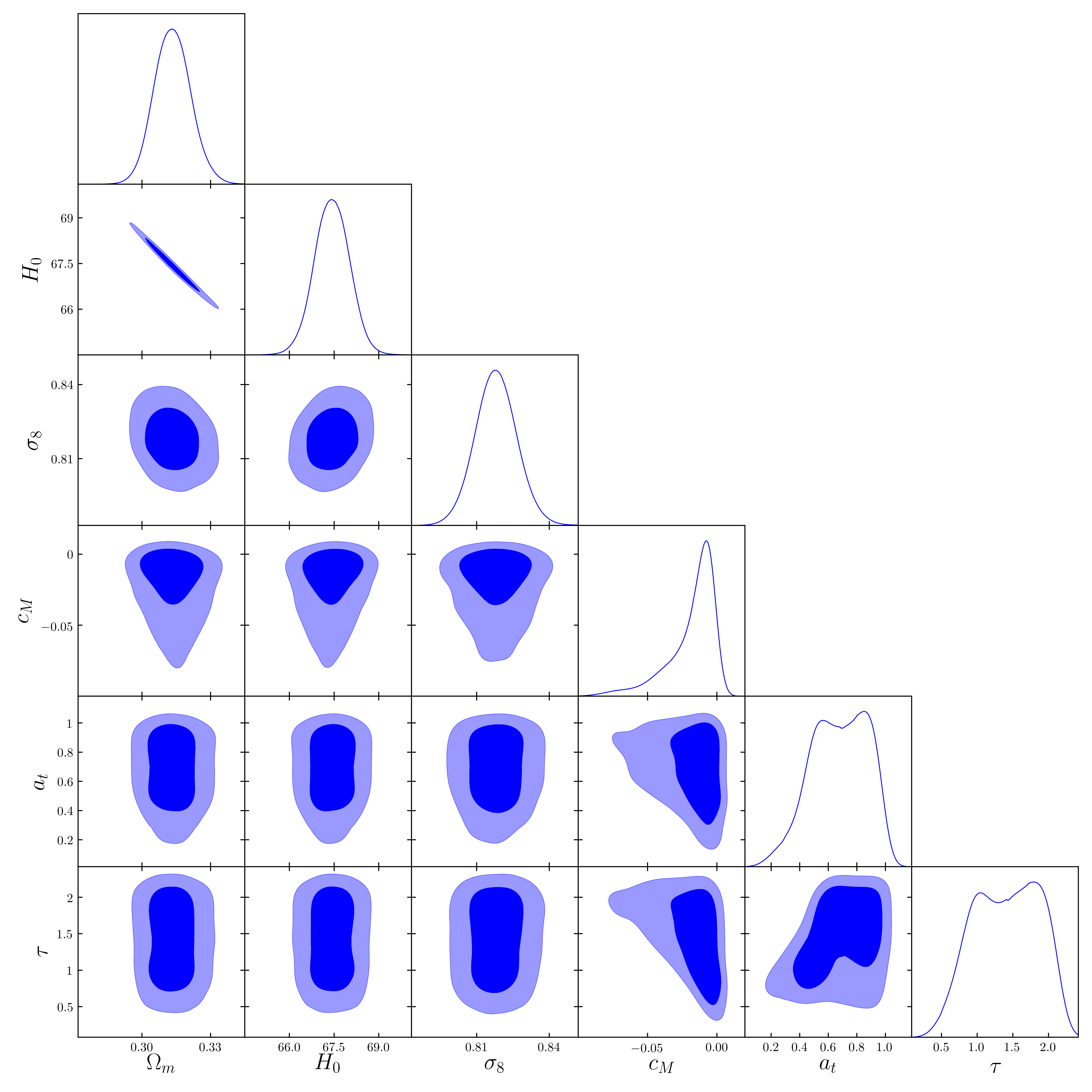} 
\caption{Triangle plot of the joint probability distributions, at 68.3\% and 95.4\% confidence levels, and marginalized one dimensional posteriors, for various cosmological and 
gravity parameters, fixing to a \lcdm\ background. 
Here $a_{t}$ and $\tau$ are free to vary. 
}
\label{fig:mcmcvarylcdm}
\end{figure*}

%%%%%%%%%%%%%%%%%%%%%%%%% 
\begin{figure}
    \centering
    \includegraphics[width=\columnwidth]{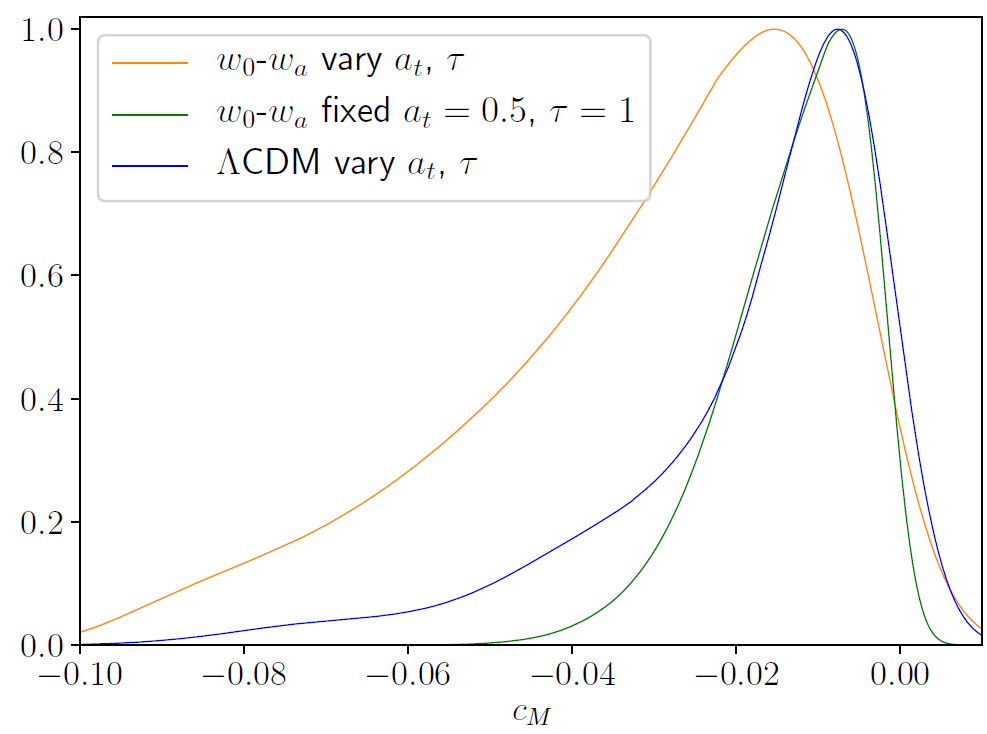}
    \caption{Marginalized one dimensional posterior comparison for the $c_{M}$ parameter between the MCMC analyses performed. We can see the shift in the 
distribution to more 
    negative values when $\tau$ and $a_{t}$ are allowed to vary. (Note the 
    tails to positive $c_M$ are artifacts of the plotting and 
    do not occur in the chains due to stability conditions.) 
    }
    \label{fig:cm_compar}
\end{figure}

For $c_M$ distinct from zero, larger amplitude in $c_M$ 
correlates with larger $\tau$. This follows from the Planck mass 
maximum being $e^{-c_{M}/\tau}$, and $\geff$ being the inverse of the Planck mass. Similarly, increasing $a_t$ moves the maximum 
deviation in $\geff$ later, decreasing its effect, and so $a_t$ 
and $c_M$ are also correlated. 

\begin{figure}
\includegraphics[width=\columnwidth]{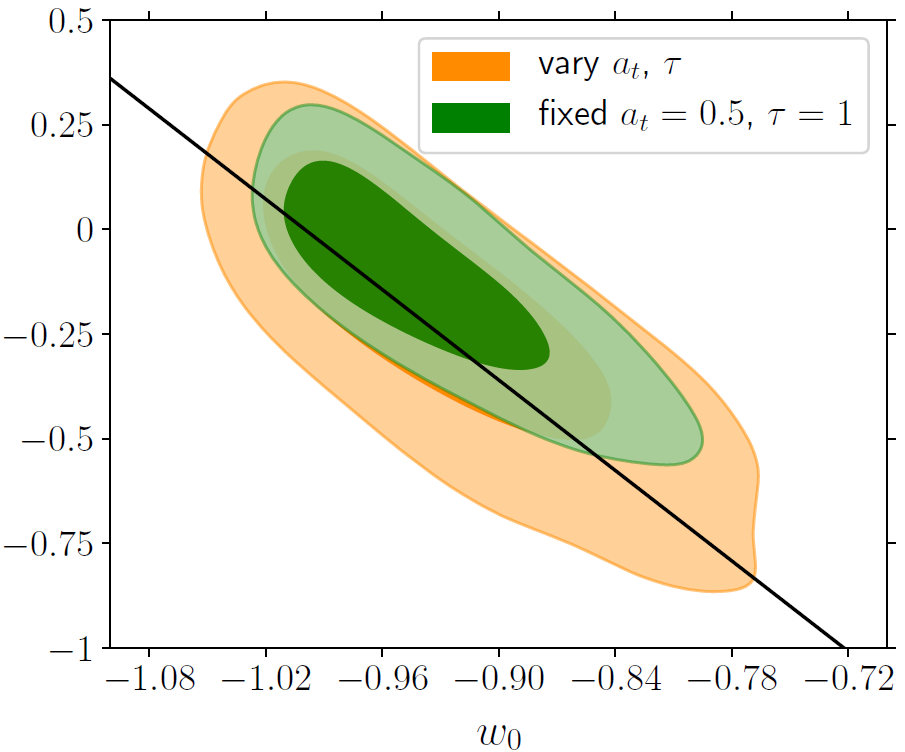}
\caption{The joint posterior between the dark energy equation of state 
parameters is shown for the two analysis cases, with the mirage line $w_a=-3.6\,(1+w_0)$ overlaid. 
}
\label{fig:w0wa_mirage}
\end{figure}

Apart from the gravity parameters, all the standard primordial cosmology parameters are consistent with the usual general relativity, 
\lcdm\ values. We list their values in the tables, but do not show them in the triangle plots in order to make the other 
parameters more visible to the reader. With regard to dark energy, note the mostly one sided distribution of $w_0$ as required by stability 
considerations. The joint posterior for $w_0$--$w_a$ shown in Fig.~\ref{fig:w0wa_mirage} 
demonstrates that mirage models come close to describing the viable models. This indicates that 
the CMB acoustic scale provides significant constraining power, and is also consistent 
with structure growth as seen in Fig.~\ref{fig:fsigw0}. The 
posterior is pulled slightly above the mirage line due to the 
BAO and supernovae which prefer a somewhat lower matter density 
at medium redshifts, and hence a more persistent dark energy ($w_0>-1$).

%%%%%%%%%%%%%%%%%%%%%%%%%%%%%%%%%%%%%%%%%%%%%%%%%%%%%%%%%%%

Finally, we then fix to the \lcdm\ background ($w_0=-1$, $w_a=0$) while 
still allowing modified gravity. 
The results are shown in Table~\ref{tab:mcmcparamvarylcdm} and in Fig.~\ref{fig:mcmcvarylcdm}. 
The standard cosmology parameters are little affected, and $\sigma_8$ still 
shows its mild suppression from the Planck GR \lcdm\ value of 0.83; the GR \lcdm\ value for the data sets we use is $\sim0.82$ (this can also be seen roughly by slicing through the 
$\sigma_8$--$c_M$ contour shown in Fig. 9 at the $c_M=0$, i.e. GR, value. For the 
modified gravity amplitude, Figure~\ref{fig:cm_compar} compares the 1D 
posteriors for $c_M$  between the three cases. They are fairly consistent 
with each other. When comparing the \lcdm\ case 
with both $w_0$--$w_a$ cases, one can see that all are consistent with 
general relativity at the 95\% confidence level. The peak of the \lcdm\ 
case is quite similar to the $w_0$--$w_a$ case with fixed $a_t$ and $\tau$, 
while like the $w_0$--$w_a$ case with varying $a_t$ and $\tau$ 
there is a tail extending to more negative $c_M$.

The $\Delta\chi^2$ between the three cases is less than 0.4, indicating 
no significant preference for either allowing the background to vary 
(note, however, that there will be regions of model space, i.e.\ $a_t$ 
and $\tau$, where a \lcdm\ background does not give a stable theory while 
a more general $w_0$--$w_a$ background does) or 
allowing $a_t$ and $\tau$ to vary. This is basically because all cases 
prefer small $c_M$ where there is less distinction between these variations.

%%%%%%%%%%%%%%%%%%%%%%%%%%%%%%%%%%%%%%%%%%%%%%
\section{Conclusions} \label{sec:concl} 

Allowing for freedom in the cosmic background history enables greater diversity 
of stable modified gravity models. In particular, for No Slip Gravity it broadens  
parameter space with $\alm<0$. To study this, we introduced a new hill-valley form 
for $\alm(a)$ that allows both increasing and decreasing Planck mass evolution. We derived 
the simple analytic form for $\ms$, and the effective gravitational strength $\geff$, plus 
analytic limits from stability considerations on some parameters ($w_0$ and $\tau$). 
Beyond No Slip Gravity we also briefly explored a generalized relation between the 
effective field theory property functions $\alb$ and $\alm$. 

For the background evolution, the dark energy mirage relation gives a reasonable 
approximation to the preferred region of effective dark energy parameter space even within 
the modified gravity theory studied. This offers a way of reducing the dimension of 
the parameter space to be fit (although we fit for the full $w_0$--$w_a$ 
space, as well as for a \lcdm\ expansion history). 

No Slip Gravity is an interesting example theory in that it has a simple relation of 
$G_{\rm matter}$ and $G_{\rm light}$ to $\ms$. Furthermore it is unusual among modified 
gravity theories in suppressing growth, as data mildly prefers. We extended previous 
analysis also to effects beyond growth, in particular $G_{\rm light}$ 
as well as $G_{\rm matter}$. 

We studied No Slip Gravity predictions for growth of large scale structure ($\fsig$), 
light propagation (decay of potentials and lensing), CMB, and ISW crosscorrelations. 
No Slip Gravity (and No Run Gravity) gives standard positive ISW-galaxy 
crosscorrelation -- as the data prefers -- unlike in some modified gravity 
models. We also found that an analytic approximation for lensing and ISW suppression 
holds for the new hill-valley model. Mirage models were demonstrated to have similar growth 
histories to each other in GR, and in modified gravity, i.e.\ mirage dark energy with 
$w_0=-0.8$ is similar to \lcdm\ even in modified gravity. This holds as well with respect 
to similar lensing suppression. 

We modified the Boltzmann code {\tt hi\_class} for this new model of No Slip Gravity (with the modified version made publicly available on GitHub at the URL give in Sec.~\ref{sec:mcmc}), 
and furthermore adapted the code to enable computation of the redshift space distortion 
observable $\fsig$ and its application in MCMC likelihood evaluation for modified gravity. 

Carrying out an MCMC analysis using current data we find the background parameters are 
consistent with general relativity and \lcdm, but the modified gravity case somewhat lowers the 
value of $\sigma_8$, even in a $\Lambda$CDM background, easing the tension with weak lensing measurements interpreted within 
GR \lcdm (taking into account the cautions of Sec. VI where the weak lensing data analysis should be done within the new theory). Note that No Slip Gravity suppresses both structure growth and lensing deflection. For the 
amplitude of the modified gravity strength, $0>c_M>-0.08$, 
i.e.\  $|\alpha_{M,{\rm max}}|<0.03$. That is, over the entire evolution the Planck mass running 
cannot be too severe and so the modified gravity 
cannot lie too far from general relativity. In addition, general relativity 
lies within the 95\% confidence level.

%%%%%%%%%%%%%%%%%%%%%%%%%%%%%% 
\section*{Acknowledgments}

We gratefully acknowledge helpful conversations with 
Miguel Zumalac{\'a}rregui. 
This work made use of the CHE cluster, managed and funded by COSMO/CBPF/MCTI, with financial support from FINEP and FAPERJ, and operating at the Javier Magnin Computing Center/CBPF. GB would like to acknowledge the State Scientific and Innovation Funding Agency of Espirito Santo (FAPES, Brazil) and the Brazilian Physical Society (SBF) through the SBF/APS PhD Exchange Program for financial support. GB would also like to thank LBL for the hospitality and financial support. FTF would like to thank the National Scientific and Technological Research Council (CNPq, Brazil) for financial support. HV would like to thank FAPES and CNPq for financial support. GB gratefully acknowledges Renan A. Oliveira and David Camarena for useful discussions in an early version of this work. 
This work is supported in part by the Energetic Cosmos Laboratory and by 
the U.S.\ Department of Energy, Office of Science, Office of High Energy 
Physics, under Award DE-SC-0007867 and contract no.\ DE-AC02-05CH11231.

\appendix 

%%%%%%%%%%%%%%%%%% 
\section{Variation of $a_t$ and $\tau$} \label{sec:apxtau} 

As described in Sec.~\ref{sec:nsg}, the values chosen for the transition time 
and width parameters, $a_t$ and $\tau$, of the hill/valley form 
for the illustrative plots were motivated 
by physical reasons of being close to the onset of cosmic acceleration and 
having the transition of order one e-fold of expansion. This also leads to an 
opportunity for the modified gravity to have an appreciable impact on 
observations. Of course in Sec.~\ref{sec:mcmc} the Monte Carlo 
analysis scans over these parameters. 

\begin{figure}[htbp!]
\includegraphics[width=\columnwidth]{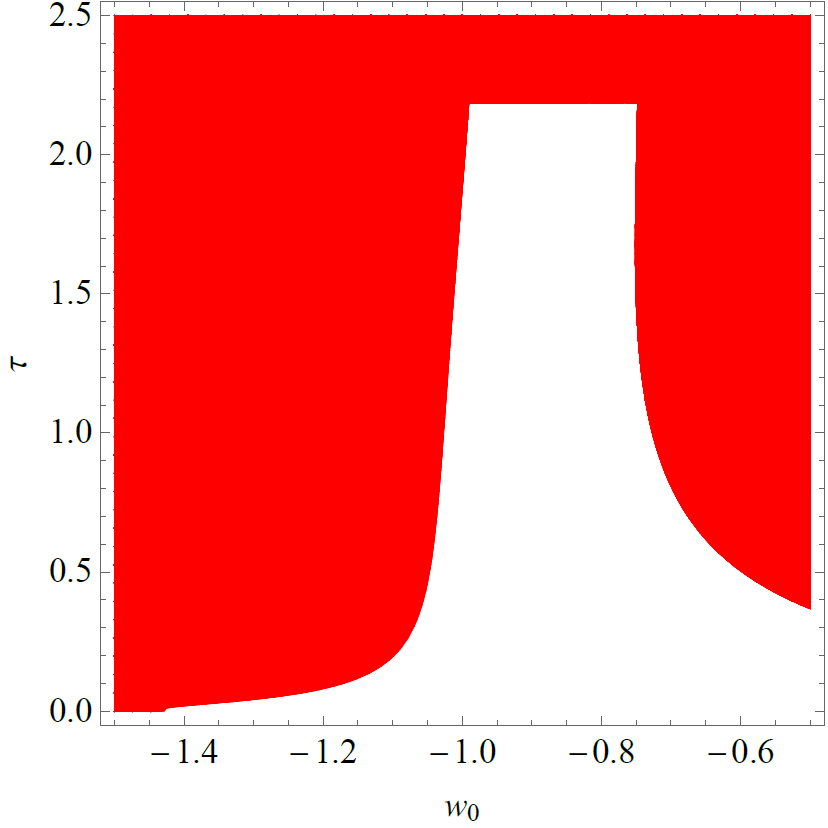} 
\caption{The stability region for the hill/valley form of $\alm$ is shown in 
the $\tau$--$w_0$ plane, for the mirage 
dark energy equation of state. The unplotted parameters are set to their 
fiducial values $c_M =-0.05$, $a_t=0.5$. } 
\label{fig:tauw0}
\end{figure}

\begin{figure}[htbp!] 
\includegraphics[width=\columnwidth]{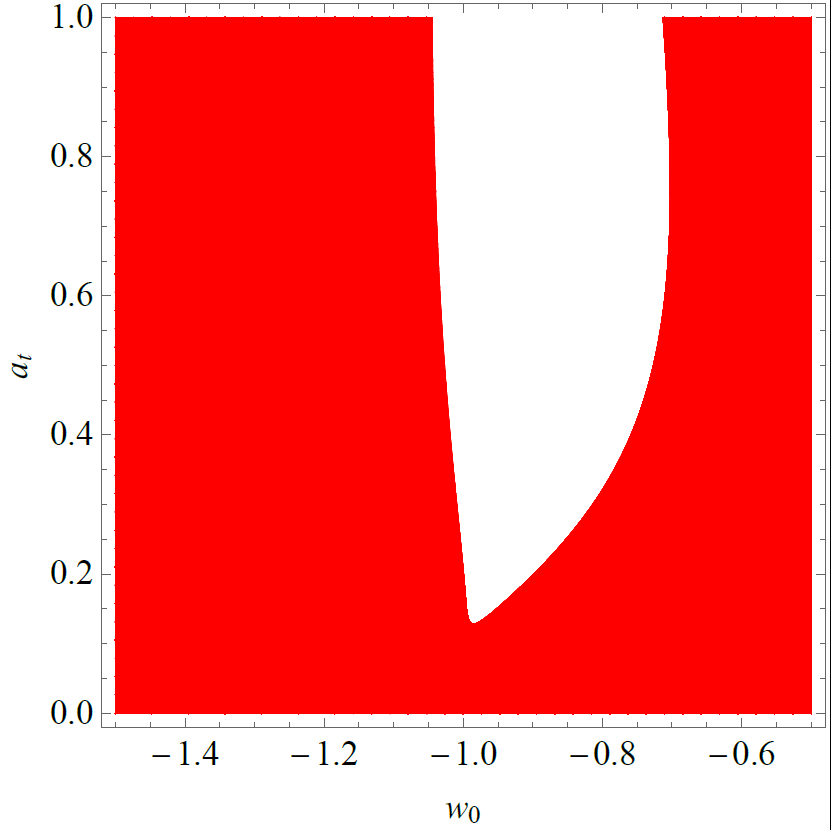}
\caption{As Fig.~\ref{fig:tauw0} but for the $a_t$--$w_0$ plane. The unplotted 
parameters are set to their fiducial values $c_M =-0.05$, $\tau=1$. }
\label{fig:atw0}
\end{figure}

Here we show that the reasonably natural values chosen, $a_t=0.5$ and $\tau=1$, 
are not special with regard to stability considerations, i.e.\ not a small 
island in parameter space. This also motivates priors for the Monte Carlo sampling. 
Figure~\ref{fig:tauw0} shows the stability region 
in the $\tau$--$w_0$ plane for the mirage model, fixing the other hill/valley 
parameters to the fiducial values: $c_M=-0.05$ and $a_t=0.5$. Values of $\tau$ 
larger than $\tau_c=(3+\sqrt{33})/4 \approx 2.19$ are ruled out by instability at 
early times (this value is independent of $c_M$ and $a_t$); the side regions are ruled out by instability at more recent times. 
Figure~\ref{fig:atw0} shows the corresponding diagram in the $a_t$--$w_0$ plane, 
with fixed $\tau=1$. A transition occurring too early gives rise to instability 
at early times.

%%%%%%%%%%%%%%%%%%%%%%%%%%%% 

%%%%%%%%%%%%%%%%%%%%%%%%%%%% 

\twocolumngrid

\end{document}